\documentstyle[prb,aps,preprint]{revtex}
\tightenlines
\begin{document}
\title{Critical properties of $S=\frac{1}{2}$ Heisenberg ladders \\
in magnetic fields}
\author{Mamoru Usami and Sei-ichiro Suga}
\address{Department of Applied Physics, Osaka University, Suita, Osaka
565-0871, Japan}
\maketitle
\begin{abstract}
The critical properties
of the $S=1/2$ Heisenberg two-leg ladders are investigated
in a magnetic field.
Combining the exact diagonalization method
and the finite-size-scaling analysis based on conformal field theory,
we calculate the critical exponents of spin correlation functions
numerically.
For a strong interchain coupling,
magnetization dependence of the critical exponents shows characteristic
behavior depending on the sign of the interchain coupling.
We also calculate the critical exponents
for the $S=1/2$ Heisenberg two-leg ladder with a diagonal interaction,
which is thought as a model Hamiltonian
of the organic spin ladder compound
%$\mbox{Cu}_2(\mbox{C}_5\mbox{H}_{12}\mbox{N}_2)_2\mbox{Cl}_4$.
$\mbox{Cu}_2(\mbox{1,4-diazacycloheptane})_2\mbox{Cl}_4$.
Numerical results are compared with experimental results
of temperature dependence of the NMR relaxation rate $1/T_1$.
\end{abstract}
\section{Introduction}
The $S=1/2$ bond alternating Heisenberg chains
with the next-nearest-neighbor interaction attract much attention.
Schematic of this model is shown in Fig.\ \ref{fig:schematic}(a).
The alternating nearest-neighbor interactions are controlled
by $J_1$ and $J_2$, and the parameter $J_3$
is the strength of the next-nearest-neighbor interaction.
Magnetic properties of the spin-Peierls compound $\mbox{CuGeO}_3$
is probably described by this model.\cite{frast}
Alternatively, as shown in Fig.\ \ref{fig:schematic}(b),
another aspect of this model is that
this model describes a spin ladder system with a diagonal interaction;
$J_3$ and $J_1$ are strength of the interaction along
the ladder and along the rung, respectively,
and the diagonal interaction is represented by $J_2$.
It is thought that
such a model describes the new spin ladder compound
$\mbox{Cu}_2(\mbox{1,4-diazacycloheptane})_2\mbox{Cl}_4$
(abbreviated CuHpCl).\cite{HR,CCLPMM,HPL}
From the theoretical point of view, the plateau on the magnetization curve
is observed in this model.\cite{TNK,totsuka}

This model is also related to the Haldane-gap system.
For $J_2=0, J_3>0$ and $-J_1/J_3 \gg 1$
or $J_3=0, J_2>0$ and $-J_1/J_2 \gg 1$,
two spins connected with the strong ferromagnetic interaction ($J_1$)
form a triplet.
Because it effectively works as a $S=1$ spin, we can regard such a model
as describing the $S=1$ Heisenberg antiferromagnetic chain
in the strong coupling limit.\cite{hida91,hida92}

Usually, such a system has a gap $\Delta$
between the singlet ground state and a triplet excited one.
An external magnetic field lowers
the energy of a triplet excited state,
but does not change the singlet ground state energy.
At a critical field $H_{c_1} (=\Delta)$, the gap disappears
and a transition from a non-magnetic state to a magnetic one occurs.
In a stronger magnetic field $H>H_{c_1}$,
the system is gapless until the magnetization is saturated,
unless the plateau appears on the magnetization curve.

One of the important features of the organic compound
is its small coupling constant.
The inorganic spin ladder compound $\mbox{SrCu}_2\mbox{O}_3$,
which is known for its excellent two-leg ladder character,
has a gap $\Delta \ge 400 \mbox{K}$.\cite{AHTIK}
Thus, the critical field $H_{c_1}$ is not accessible.
On the other hand, the organic compound CuHpCl
which is thought to be the $S=1/2$ Heisenberg two-leg ladder compound,
has a smaller gap $\Delta \sim 10.5 \mbox{K}$.\cite{chabo}
Because the critical field $H_{c_1} \sim 7.5 \mbox{T}$ can be accessed,
some experiments\cite{HR,CCLPMM,chabo,HRBT,CJFLBHP}
have been done in the magnetic state which has finite magnetization.

On the basis of bosonization,
Chitra and Giamarchi \cite{CG} have studied such a gapped system,
at the critical field $H_{c_1}$ and in its vicinity.
They have concluded that for the $S=1/2$ Heisenberg two-leg ladder system,
the critical exponent does not depend on a magnetic field
in the region close to $H_{c_1}$.
Nevertheless, this is inconsistent with experimental data
observed in CuHpCl.\cite{chabo}
Since CuHpCl has a strong interchain coupling,
it is important to study a spin ladder system
with a strong interchain coupling theoretically.

In this paper,
we will investigate the critical properties of
the spin ladder system with or without a diagonal interaction
(see Fig.\ \ref{fig:schematic}(b))
in the magnetic state.
In Sec.\ \ref{sec:numerical}, we present a model Hamiltonian
and the numerical method\cite{ST,ST2} developed by Sakai and Takahashi.
Using this method,
we calculate the critical exponent of the spin correlation function
in the magnetic state for the $S=1/2$ Heisenberg two-leg ladder with
the strong antiferromagnetic interchain interaction in Sec.\ \ref{sec:AF},
and with the strong ferromagnetic one in Sec.\ \ref{sec:F}.
Finally, we calculate the critical exponents
taking a diagonal interaction into account,
and compare the numerical results with experimental data\cite{chabo}
observed in CuHpCl in Sec.\ \ref{sec:diag}.
The last section is devoted to our conclusion.

\section{Numerical Method}
\label{sec:numerical}
We consider the following Hamiltonian,
\begin{eqnarray}
{\cal H} &=& {\cal H}_0 + {\cal H}_1  \label{eqn:hamiltonian}, \\
{\cal H}_0 &=& 2 \sum_{i=1}^{N/2} 
\{ J_1 \mbox{\boldmath $S$}_{2i-1} \cdot \mbox{\boldmath $S$}_{2i} 
+ J_2 \mbox{\boldmath $S$}_{2i} \cdot \mbox{\boldmath $S$}_{2i+1} \}
+ 2 J_3 \sum_{i=1}^N  \mbox{\boldmath $S$}_i \cdot \mbox{\boldmath
$S$}_{i+2}
\label{eqn:hamiltonian0}, \\
{\cal H}_1 &=& -g\mu_BH \sum_{i=1}^N  S_i^z,
\end{eqnarray}
where $\mbox{\boldmath $S$}_i$ denotes the $S=1/2$ operator in the
$i$-th site,
$N$ is the total number of sites,
and $H$ is the strength of a magnetic field applied along $z$-axis.
The distance between unit cells
which consist of neighboring two sites
is set equal to unity.
Afterward we set $J_3=1$ and $g \mu_B=1$.
The periodic boundary condition is applied.
Since the system has a rotational symmetry about $z$-axis
and a translational symmetry,
we can classify the Hamiltonian into the subspace
according to the magnetization $M=\sum S_i^z$ and the wave vector $k$.
The lowest energy in each subspace is calculated
using the Lanczos algorithm.
For the $N$-site system,
we define $E_k(N,M)$ as the lowest energy of ${\cal H}_0$ in the subspace
specified by the magnetization $M$ and the wave vector $k$.
For given $N$ and $M$,
$E_k(N,M)$ takes the minimum at $k=k_0$.
From now on, we simply describe $E_{k_0}(N,M)$ as $E(N,M)$.
Note that $k_0=0$ for even $M$ and $k_0=\pi$ for odd $M$,
for the parameters used in this work.

To study critical properties of the system in a massless region,
we apply the numerical method\cite{ST,ST2} developed by Sakai and Takahashi,
which is summarized below.
First, we consider the ground state energy.
According to the conformal field theory
for the one-dimensional quantum system,\cite{cft}
the ground state energy of the massless system depends on the system
size as
\begin{equation}
\frac{1}{N} E(N,M) \sim \varepsilon (m) - A \frac{1}{N^2}
\;\;\;(N \rightarrow \infty),
\label{eqn:size}
\end{equation}
where $\varepsilon(m)$ is the ground state energy per site
in the thermodynamic limit,
$m=M/N$ is the magnetization per site,
$v_s$ is the sound velocity,
and $A=\frac{\pi}{3} c v_s $ with $c$ being the central charge.
We approximate $v_s$ numerically to be,
\begin{equation}
v_s = \lim_{k \rightarrow +k_0} \frac{dE_k}{dk} \sim
\frac{1}{|k_1-k_0|} \left( E_{k_1}(N,M)-E(N,M) \right),
\label{eq:velocity}
\end{equation}
where $k_1$ is the wave vector closest to $k_0$,
i.e. $|k_1-k_0|=\frac{4\pi}{N}$.
Using those values of $A$ and $v_s$, we can obtain the central charge $c$.

Next, we consider some excitations
to obtain the asymptotic form of the spin correlation function
in the magnetic state.
We define $\delta k$ as the difference of the wave vector
between the ground state and the excited one.
The spin excitation which increases the magnetization
($M \rightarrow M+1$) has $\delta k=k_c$.
In this paper, $k_c$ always equals $\pi$.
Moreover, in the magnetic state,
a gapless excitation can exist at the soft mode $\delta k=2k_F$.
Because the system is gapless in the magnetic state,
the spin correlation function should decay algebraically
in the ground state.
We define a spin operator $\phi_i$
as a combination of the two $S=1/2$ operators
$\mbox{\boldmath{$S$}}_{2i}$ and $\mbox{\boldmath{$S$}}_{2i+1}$.
The asymptotic forms of the correlation function
of the new spin operator $\phi$ should be\cite{sakai}
\begin{eqnarray}
\langle \phi_0^x \phi_r^x \rangle      &\sim& \cos (k_c r) r^{-\eta}
\;\;\;(r \rightarrow \infty), \\
\langle \phi_0^z \phi_r^z \rangle - 4m^2 &\sim& \cos (2k_Fr) r^{-\eta^z}
\;\;\;(r \rightarrow \infty).
\end{eqnarray}
According to the conformal field theory,\cite{cft,cft2}
the critical exponents $\eta$ and $\eta^z$ are obtained as\cite{ST2}
\begin{eqnarray}
\eta(N)   & = & \frac{E(N,M+1)+E(N,M-1)-2E(N,M)}{E_{k_1}(N,M)-E(N,M)},
\label{eqn:eta} \\
\eta^z(N) & = & 2 \frac{E_{2k_F}(N,M)-E(N,M)}{E_{k_1}(N,M)-E(N,M)}.
\label{eqn:etaz}
\end{eqnarray}
In the following sections, we calculate the critical exponents
using Eqs.\ (\ref{eqn:eta}) and (\ref{eqn:etaz}).

\section{Strong Antiferromagnetic Interchain Interactions}
\label{sec:AF}
We study the $S=1/2$ Heisenberg two-leg ladder system
with a strong antiferromagnetic interchain interaction.
The parameters we use are  $J_1=5.0$ and $J_2=0$.
At first, to examine whether Eq.\ (\ref{eqn:size}) is satisfied,
we calculate size dependence of the ground state energy.
In Fig.\ \ref{fig:size}, we plot the ground state energy $E(N,M)/N$ as
a function of $1/N^2$
for $m=1/8, 1/4$ and $3/8$.
Because these lines seem linear,
we conclude that Eq.\ (\ref{eqn:size}) is satisfied.
For each $m$,
the $y$-intercept of this line is the ground state energy
in the thermodynamic limit
and the gradient of this line is $-A$.
From the values of $A$ and $v_s$,
we can estimate the central charge; $c=\frac{3}{\pi}A/v_s$.
The estimated value of the central charge
is equal to unity within a few percent.
Therefore, the system belongs to the universality class
of the Tomonaga-Luttinger liquid.

We also calculate the critical exponents $\eta$ and $\eta^z$
by Eqs.\ (\ref{eqn:eta}) and (\ref{eqn:etaz}).
In Fig.\ \ref{fig:eta_finite}, 
$\eta(N)$ and $\eta^z(N)$ are shown as a function of the magnetization.
Because $\eta(N)$ and $\eta^z(N)$ obviously depend on the system size,
we plot $\eta(N)$ and $\eta^z(N)$ versus $N^{-2}$ for 
$m=1/4, 3/8$ in Fig.\ \ref{fig:gaiso}.
It seems that $\eta(N)$ and $\eta^z(N)$ are proportional to $N^{-2}$
except $\eta^z$ for $m=1/4$.
We evaluate the values of $\eta$ and $\eta^z$ in the thermodynamic limit
as the $y$-intercept of the lines
obtained by the least-squares method, except $\eta^z$ for $m=1/4$.
The deviation from the line is so small (less than $0.1\%$)
that we neglect it.
For $\eta^z$ for $m=1/4$,
we extrapolate using the only two values for the largest system and
for the next-largest one.
For other $m$,
although $\eta(N)$ and $\eta^z(N)$ can be calculated
only for two or three values of $N$,
we presume that $\eta(N)$ and $\eta^z(N)$ are proportional to $N^{-2}$.
In Fig.\ \ref{fig:eta_af}, $\eta$ and $\eta^z$ in the thermodynamic
limit are shown.

If the system is described by the Tomonaga-Luttinger liquid,
the universal relation $\eta \cdot \eta^z=1$ must be satisfied.\cite{LP}
The values of $\eta \cdot \eta^z$ in the thermodynamic limit
are shown in Table \ref{tab:central_af}.
From these values,
we conclude that the relation $\eta \cdot \eta^z=1$ is satisfied well.
This is consistent with $c=1$ obtained above.
We note that the value of $\eta \cdot \eta^z$ for $m=1/4$ is slightly smaller
than that for other $m$.
This may be due to the peculiar size dependence of $\eta^z(N)$ for $m=1/4$.
At $m=1/4$, the plateau exists on the magnetization curve
when the coupling constants are selected adequately.\cite{TNK}
This means the system has a gap and Eq.\ (\ref{eqn:size}) is not satisfied.
Nevertheless, for the parameter we use here,
there must not be the plateau on the magnetization curve at $m=1/4$.
To declare the relation between
the size dependence of $\eta^z$ for $m=1/4$
and the plateau on the magnetization curve,
further studies are needed.

\section{Strong Ferromagnetic Interchain Interactions}
\label{sec:F}
Next, we consider critical exponents
for the spin ladder with a strong ferromagnetic interchain interaction,
using $J_1=-5.0$ and $J_2=0$.
This system is expected to behave like the Haldane-gap system,\cite{hida91}
because the strong interchain coupling makes two spins triplet along a rung
and it works as one $S=1$ spin effectively.
Since we cannot obtain the value of $E_{2k_F}$ accurately,
we calculate only $\eta$ in the magnetic state.
The magnetization dependence of $\eta$ is shown in Fig.\ \ref{fig:eta_f}.
Because the size dependence of $\eta$ is small,
we neglect the size correction.
This result is in good agreement with the value of $\eta$ for $S=1$
antiferromagnetic chain\cite{ST}
not only qualitatively but quantitatively.

As mentioned above, we cannot calculate $\eta^z$ directly.
However, if the system is described by the Tomonaga-Luttinger liquid
and the central charge $c$ is equal to unity,
$\eta \cdot \eta^z =1$ must be satisfied also in this system.
Thus, we can estimate $\eta^z=\eta^{-1}$.
To ascertain $c=1$,
we calculate size dependence of the central charge
and extrapolate the result to the thermodynamic limit.
For the $N$-site system,
we define $A(N)$ and $v_s(N)$ as follows.
Let $N'$ be the greatest integer less than $N$
so that $m=M/N=M'/N'$.
$E(N,M)/N$ and $E(N',M')/N'$ are calculated numerically.
Fitting these two values with Eq.\ (\ref{eqn:size}),
we obtain $A(N)$ as the gradient of this line.
We also define $v_s(N)$ as $v_s$ for finite $N$,
using Eq.\ (\ref{eq:velocity}).
Consequently, the central charge of the $N$-site system is defined as
$c(N)=\frac{3}{\pi}A(N)/v_s(N)$.
For $m=1/4$ and $3/8$,
the size dependence of $c(N)$ is shown in Fig.\ \ref{fig:central}.
Since $c(N)$ seems to be proportional to $N^{-2}$,
we fit these data with a straight line using the least-squares method
and define $c$ as the $y$-intercept of this line.
The results are $c=0.995$ for $m=1/4$ and $c=0.998$ for $m=3/8$.
These values suggest that $c=1$.
For the $S=1$ antiferromagnetic chain with finite magnetization ($0<m<1$),
the central charge is unity
and the relation $\eta \cdot \eta^z =1$ is satisfied.\cite{ST}
Since the value of $\eta$ obtained here is in good agreement
with that for the $S=1$ antiferromagnetic chain,
the magnetization dependence of $\eta^z$ in this system
must have similar behavior to that for the $S=1$ antiferromagnetic chain.

Comparing these two results
(Fig.\ \ref{fig:eta_af} and Fig.\ \ref{fig:eta_f}),
magnetization dependence of critical exponents
shows characteristic behavior depending on the sign of
the interchain interaction $J_1$.
Similar behavior was found by Sakai \cite{sakai}
for the $S=1/2$ bond-alternating chains,
which is described by the Hamiltonian (\ref{eqn:hamiltonian})
restricting $J_3$ to zero and $J_1$ to unity.
He has concluded that
this system is described by the Tomonaga-Luttinger liquid
and its critical exponents satisfy
$\eta^z/2 < 1 < 2\eta$ for $J_2 > 0$
and $2\eta < 1 < \eta^z/2$ for $J_2 <0$, in the magnetic state.
Furthermore,
$2\eta=\eta^z/2=1$ in the limits of $m \rightarrow 0+$ and
$m \rightarrow 1/2-$,
regardless of the sign of $J_2$.

\section{Spin Ladder Material:
$\mbox{C\lowercase{u}}_2(\mbox{\lowercase{1,4-diazacycloheptane}})_2
\mbox{C\lowercase{l}}_4$}
\label{sec:diag}
As mentioned previously, CuHpCl
is described well by the isolated coupled $S=1/2$ chain so far.
Recently, Chaboussant et al.\cite{chabo} have presented
NMR study of this material in magnetic fields.
They have measured the proton NMR relaxation rate $1/T_1$,
varying magnetic fields.
In a magnetic state, they have observed the divergence of $1/T_1$
at low temperature.
If the system is gapless,
$1/T_1$ diverges algebraically at low temperature,
\begin{equation}
\frac{1}{T_1} \propto T^{-\alpha}.
\end{equation}
They have concluded that 
$\alpha$ equals $1/2$ at $m\rightarrow 0+$ and
increases with magnetic fields till certain strength of magnetic fields.
After that, an increase in the magnetic field weakens the divergence.
Anyway, $\alpha$ increases
with the magnetic field near $H_{c_1}$ (for small $m$).
As they have pointed out,
the exponent $\alpha$ is related to the exponent $\eta$
as $\eta = 1-\alpha$.
Hence, $\eta$ must decrease with an increase
in the magnetic field for small $m$.

It is believed that this material is a model system in the
strong coupling ($J_1/J_3 \sim 5.5$).\cite{CJFLBHP}
Thus, the behavior of $\eta$ must be similar to that of
the spin ladder system with a strong antiferromagnetic interchain
interaction as shown in Sec.\ \ref{sec:AF}.
However, in Fig.\ \ref{fig:eta_af}, the exponent $\eta$ increases
with an increase in  $H$ from $H_{c_1}$.
This means that our numerical results differ from the experimental findings.

It is thought that CuHpCl
has not only ordinary ladder interactions $J_1$ and $J_3$
but a diagonal interaction $J_2$.\cite{HR,CCLPMM,HPL}
To decide coupling constants,
Hayward et al.\cite{HPL} have calculated magnetization curves and
fitted them on the experimental data.
They have suggested that the diagonal interaction is ferromagnetic
and its value is $J_2/J_1=-0.1$ for $J_3/J_1=0.18$.
Hence, we calculate $\eta$ and $\eta^z$ again, turning $J_2$ on.
The parameters we use are $J_1=5.5$ and $J_2=-0.55$,
corresponding to the above ratio.
The values of $\eta(N)$ and $\eta^z(N)$
obtained from Eqs.\ (\ref{eqn:eta}) and (\ref{eqn:etaz})
have size dependence.
Thus, we extrapolate to the thermodynamic limit
by the same way in Sec.\ \ref{sec:AF}.
Note that all of the $\eta(N)$ and $\eta^z(N)$,
including $\eta^z$ for $m=1/4$,
are proportional to $N^{-2}$ when $m$ is fixed.
The numerical results are shown in Fig.\ \ref{fig:eta_diag}
and the values of $\eta \cdot \eta^z$ are shown
in Table \ref{tab:central_diag}.
Also in this system, the relation $\eta \cdot \eta^z =1$ is satisfied well.

In Fig.\ \ref{fig:eta_diag},
the maximum (minimum) of $\eta$ ($\eta^z$) is smaller (larger)
than the results in Sec.\ \ref{sec:AF}.
Further, the value of magnetization at which $\eta$ takes the maximum
shifts to a larger value region.
Nevertheless, the ferromagnetic diagonal interaction does not change
the magnetization dependence of $\eta$ and $\eta^z$, qualitatively.
Our calculation still differs from experimental results.

On actual materials, more complicated mechanisms may exist,
e.g. interladder coupling, lattice distortion and so on.
Recently, Nagaosa and Murakami\cite{NM} have treated
the $S=1/2$ Heisenberg two-leg ladder system
taking a lattice distortion into account.
They have concluded that the lattice will be distorted to modulate the
interchain coupling with the incommensurate wave vector
which is proportional to the magnetization.
To explain experimental results, such effects may be important.
Moreover, the coupling constants are still controversial.\cite{HRBT}
To determine the coupling constants,
the value of magnetization at which $\eta$ takes the maximum may be useful,
because the experimental determination of the absolute value of
the critical exponent is formidable.

\section{Conclusion}
We have calculated the critical exponents of the spin correlation functions
of the $S=1/2$ Heisenberg two-leg ladder system with a diagonal interaction
in a magnetic field.
First, we have considered pure spin ladders
(a diagonal interaction is set to zero).
Depending on the sign of the interchain coupling,
the magnetization dependence of the critical exponents
exhibits characteristic behavior.
For a ladder with a strong ferromagnetic interchain interaction,
the critical exponents depend on the magnetization
like that for the $S=1$ antiferromagnetic chains
not only qualitatively but quantitatively.
Next, we have taken a ferromagnetic diagonal interaction into account
in connection with CuHpCl.
This diagonal interaction makes the maximum of $\eta$ small
and the minimum of $\eta^z$ large.
Also, the magnetization at which $\eta$ ($\eta^z$) takes
the maximum (minimum) becomes large.
Comparing these results with experimental data,
there is qualitative difference.
To explain experimental results,
more complicated effects may be important.

\section*{Acknowledgement}

We would like to thank G. Chaboussant for sending us their manuscript
prior to publication and for fruitful discussions.
Our computational programs are based on TITPACK Ver. 2 by H. Nishimori.
Most of numerical computation in this work was supported
by the Yukawa Institute for Theoretical Physics.
This work was supported by the Grants-in-aid No. 10640344 for
Scientific Research from the Ministry of Education, Science and Culture, Japan.

\begin{figure}
\caption{Schematics of the model considered in this paper.}
\label{fig:schematic}
\end{figure}

\begin{figure}
\caption{
Plots of $E(N,M)/N$ vs. $1/N^2$ with $m$ fixed
for $1/8$, $1/4$ and $3/8$.
The origin is shifted along the vertical axis.
The values of points $A$ and $B$ are as follows:
$A=-2.82$, $-1.56$, $0.0$;
$B=-2.77$, $-1.51$, $0.05$
for $m=1/8$, $1/4$ and $3/8$, respectively.
The dashed lines are to guide the reader's eye.
}
\label{fig:size}
\end{figure}

\begin{figure}
\caption{
Magnetization dependence of the critical exponents
$\eta$ and $\eta^z$ for various $N$.
The parameters used here are $J_1=5.0$ and $J_2=0$.
The dashed lines are to guide the reader's eye.
}
\label{fig:eta_finite}
\end{figure}

\begin{figure}
\caption{
Size dependence of $\eta$ (upper two figures)
and $\eta^z$ (lower two figures) with $m$ fixed.
The horizontal axis means $N^{-2}$.
The dashed lines are obtained using the least-squares method
and the dotted line is to guide reader's eyes.}
\label{fig:gaiso}
\end{figure}

\begin{figure}
\caption{
Magnetization dependence of the critical exponents $\eta$ and $\eta^z$
for the $S=1/2$ two-leg ladder
with a strong antiferromagnetic interchain interaction
($J_1=5.0, J_2=0$) in the thermodynamic limit.
The dashed lines are to guide the reader's eye.
}
\label{fig:eta_af}
\end{figure}

\begin{figure}
\caption{
Magnetization dependence of the critical exponent $\eta$
for the $S=1/2$ two-leg ladder
with a strong ferromagnetic interchain interaction ($J_1=-5.0, J_2=0$).
The dashed lines are to guide the reader's eye.
}
\label{fig:eta_f}
\end{figure}

\begin{figure}
\caption{Size dependence of the central charge $c$.
The dashed lines are obtained using the least-squares method.
}
\label{fig:central}
\end{figure}

\begin{figure}
\caption{
Magnetization dependence of the critical exponents $\eta$ and $\eta^z$
for the $S=1/2$ two-leg ladder with a diagonal interaction
($J_1=5.5, J_2=-0.55$).
The results for some $N$ and extrapolated values are shown.
The dashed and solid lines are to guide the reader's eye.
}
\label{fig:eta_diag}
\end{figure}

\begin{table}
\caption{
The values of $\eta \cdot \eta^z$ in the thermodynamic limit
for the $S=1/2$ two-leg ladder with a strong antiferromagnetic interchain
interaction
($J_1=5.0, J_2=0$).}
\label{tab:central_af}
\begin{tabular}{cccccccccc}
$m$ & 1/12 & 1/8 & 1/6 & 1/4 & 1/3 & 3/8 & 2/5 & 5/12 & 7/16 \\
\hline
$\eta \cdot \eta^z$ & .993 & .998 & .997 & .977 & .999 &
 .999 & .999 & .999 & 1.000 \\
\end{tabular}
\end{table}

\begin{table}
\caption{
The values of $\eta \cdot \eta^z$ in the thermodynamic limit
for the $S=1/2$ two-leg ladder with a diagonal interaction
($J_1=5.5, J_2=-0.55$).
}
\label{tab:central_diag}
\begin{tabular}{cccccccccc}
$m$ & 1/12 & 1/8 & 1/6 & 1/4 & 1/3 & 3/8 & 2/5 & 5/12 & 7/16 \\
\hline
$\eta \cdot \eta^z$ & .996 & .999 & .999 & .995 & 1.000 &
 1.000 & 1.000 & 1.000 & 1.000 \\
\end{tabular}
\end{table}

\end{document}